\input harvmac

\lref\strov{A. Strominger and C. Vafa, {\it
Microscopic Origin of the Bekenstein-Hawking Entropy},
{\tt hep-th/9601029}.}

\lref\mal{J. Maldacena and A. Strominger,
{\it Statistical Entropy of De Sitter Space},
{\tt gr-qc/9801096}.}

\lref\malsthaw{J. Maldacena, S. Hawking,  and A. Strominger,
{\it DeSitter entropy, Quantum Entanglement and AdS/CFT},
{\tt hep-th/0002145}.
}

\lref\ban{T. Banks,
{\it Cosmological M Theory, SUSY Breaking, 
and the Cosmological Constant},
 talk at Strings'00.}

\lref\witt{E. Witten, 
{\it Quantum Gravity in deSitter space},
talk at Strings'01.}

\lref\ber{F. A. Berezin, {\it  Quantization},
Izvestiya AN USSR, ser.math. 38 (1974) 1116.}

\lref\bere{F. A. Berezin,
{\it General Concept of Quantization}, 
Comm. Math. Phys. 40 (1975) 153.}

\lref\beres{F. A. Berezin,
{\it Quantization in Complex Symmetric Spaces},
Izvestiya AN USSR, ser. math. 39 (1975)363.}

\lref\gibhaw{G.W. Gibbons and S.W. Hawking, 
Phys. Rev.D15(1977)2752.}

\lref\bsv{R. Britto-Pacumio, A. Strominger, and A. Volovich,
{\it Holography for Coset Spaces},
{\tt hep-th/9905211}.}

\lref\hor{V. Balasubramanian, P. Horava and D. Minic, 
{\it DeSitter Entropy and String Theory},
work in progress, as reported by P. Horava in a talk at 
Strings'01.}
 
\lref\membr{V.P. Frolov and I.D.  Novikov, 
{\it Black Hole Physics:
Basic Concepts and New Developments,}
Kluwer Academic Publ. (1998).}

\lref\seiw{N. Seiberg and E. Witten,
{\it String Theory and Noncommutative Geometry},
{\tt hep-th/9908142}.}

\lref\kon{ M. Kontsevich,
{\it Deformation Quantization of Poisson Manifolds},
{\tt q-alg/9709040}.}

\lref\bousso{R. Bousso,
{\it Positive Vacuum Energy and the N-bound},
{\tt hep-th/0010252}.
}

\def\kh{K\"{a}hler }

\Title{\vbox{\baselineskip12pt\hbox{}}}
{\vbox{
\centerline{Discreteness in deSitter Space and}
\smallskip
\smallskip
\centerline{Quantization of K\"{a}hler Manifolds}}}
\centerline{Anastasia Volovich
\foot{{\tt nastya@physics.harvard.edu}}
\foot{On leave from L. D. Landau Institute for Theoretical Physics,
Moscow, Russia}
}
\bigskip\centerline{Department of Physics}
\centerline{Harvard University}
\centerline{Cambridge, MA 02138}
\vskip .3in \centerline{\bf Abstract}

\smallskip

Recently, it has been proposed  
that the dimension of the Hilbert space of quantum
gravity in deSitter space is finite and moreover
it is expressed in terms of the coupling constants by using
the entropy formula. A weaker conjecture would be that
the coupling constant in deSitter space should take only discrete
values not necessarily given by the entropy formula. 
We discuss quantization of the horizon in deSitter space 
by using Berezin's functorial 
quantization of \kh manifolds and argue that the 
weak conjecture is valid
for Euclidean deSitter space. 
Moreover it can be
valid for a class of bounded complex symmetric spaces.

\Date{January 2001}

\newsec{Introduction}

An important problem is to extend the microscopic
derivation \strov\ of the Bekenstein-Hawking formula for
the black hole entropy to the case of the deSitter horizon.
In three dimensions this problem has been solved by Maldacena
and Strominger \mal\ by using the Chern-Simons approach,
see also \malsthaw .
 
Recently, it has been proposed by Banks \ban\ and 
has been discussed by Witten \witt\
that the dimension of the Hilbert space of quantum
gravity in deSitter space might be finite and moreover
it is expressed in terms of the coupling constants by using
the formula for the entropy. A weaker conjecture would be that
the coupling constant in deSitter space should take only discrete
values not necessarily given by the entropy formula. 

To explore the weak conjecture we discuss quantum
mechanics of the four dimensional deSitter ($dS_4$) space. 
The horizon in the 
Euclidean $dS_4$ space
is the two dimensional sphere.
By using Berezin's functorial 
quantization of \kh manifolds \ber, \bere, \beres\
 we argue that the 
weak conjecture might be valid
for Euclidean deSitter space. Moreover it can be
valid for a class of bounded complex symmetric spaces.

\newsec{DeSitter Entropy and Quantum Theory}
 
DeSitter spacetime is the maximally symmetric solution
of the Einstein equations with positive cosmological constant
$\Lambda$. In spherically symmetric coordinates
the metric has the form
\eqn\metr{ds^2=-(1-\Lambda r^2/3)dt^2+dr^2/(1-\Lambda r^2/3)
+r^2d\Omega^2.}
The surface $r^2=3/\Lambda$ is the horizon,
which has entropy \gibhaw
\eqn\ent{S=3\pi /\hbar G_N \Lambda.}
We will keep the Planck constant and
set  $G_N =1,~\Lambda=3$. Euclidean de Sitter space
is the sphere $S^4$ and the entropy is
\eqn\entr{S=\pi /\hbar.}
Banks \ban\ has argued that the Hilbert space of
quantum gravity in asymptotically
deSitter spacetime has a finite dimension $N$ and
\eqn\entro{S=\ln N.}
Recently, this proposal has been discussed by Witten
\witt\ and Balasubramanian,  Horava, Minic \hor.
A bound on matter entropy 
in asymptotically deSitter spaces
supporting the proposal was obtained by Bousso
\bousso.
If this proposal is valid then from \entr\ and \entro\
we obtain that the Planck constant should take
only a discrete set of values.
If the coupling constant 
(the Planck constant in our setting)
takes only a discrete set of values not necessarily
consistent with \entr\ and \entro\ 
then we will talk about 
the discreteness of dS in the weak sense.

According to the membrane paradigm \membr\  all the 
properties of black holes can be derived from the 
consideration of the horizon as a physical membrane.
Therefore if one considers quantum gravity then
it seems natural to quantize the horizon.

In the Euclidean formulation the spacetime is $S^4$
and the horizon is the two dimensional sphere $S^2$
embedded into $S^4$. To understand the proposal
we consider quantum mechanics of the horizon, i.e.
quantization of $S^2$. Quantization of $S^2$ has
been performed by Berezin (see next section) 
and it is quite remarkable that
he has found that the Hilbert space of the corresponding
quantum mechanics has a finite dimension and
that the Planck constant should 
take only discrete set of values. Therefore
one has a confirmation of the conjecture on the
discreteness of dS in the weak sense.

\newsec{Berezin's Quantization of \kh Manifolds}

Berezin \ber\ has developed a general approach to quantize
\kh manifolds. Berezin's quantization possesses the natural
functorial properties. A remarkable property of this quantization
is that the Hilbert space of quantum mechanics
of a bounded complex
symmetric space has a finite dimension and
the Planck constant takes only discrete values.
If one makes a rescaling to set the Planck constant equal to
1 then an appropriate combination
of the coupling constants will take discrete values. 
These properties support the weak conjecture.

Let $(M,\omega)$ be a Poisson manifold, where $M$
is a manifold and $\omega$ is a skew-symmetric tensor
field. Let us denote by $A(M)$ the Lie algebra of differentiable
functions on $M$ with the Poisson bracket
\eqn\poi{\{f,g\}=\omega^{ik}{\partial f \over \partial x^i}
{\partial g \over \partial x^k}.}
{\it Quantization} of $A(M)$ is defined as a family
$\{A_{\hbar}\}$ of associative algebras where 
the index $\hbar$ runs through
a subset of positive real numbers. The algebra $A_{\hbar}$
consists of the functions on $M$ with the multiplication
$f*g$\foot{Here $f*g$ should be a function but
not just a formal series in $\hbar$ as in deformation
quantization \kon.}, with the following 
properties
\eqn\corr{\lim_{\hbar\to 0}f*g=fg,~~~~~
\lim_{\hbar\to 0}{1\over \hbar}(f*g-g*f)
=i\{f,g\}.}
See \ber\  for further details. 
Such quantization has been performed for \kh manifolds.
Let $M$ be an $n$ dimensional \kh 
with metric $ds^2=g_{i\bar{k}}dz^idz^{\bar{ k}}$
and symplectic form $\omega=g_{i{\bar k}}dz^i \wedge dz^{{\bar k}}$,
$d\omega =0$. Let $F(z,{\bar z})$ be the \kh potential,
$g_{i{\bar k}}=-{\partial^2 \over {\partial z^i\partial z^{\bar k}}}
\ln F.$
Now let us consider the Hilbert space $H_\hbar$ of 
analytic functions on a submanifold
${\tilde M}$ of $M$ with the scalar product
\eqn\scal{(f,g)=c(\hbar)\int f(z){\bar g(z)}F^{1/\hbar}(z,{\bar z})
d\mu(z,{\bar z}),}
where $d\mu$ is the measure 
$d\mu=\omega^n$
and
\eqn\cish{c^{-1}(\hbar)=\int F^{1/\hbar}(z,{\bar z})d\mu(z,{\bar z}).}
Take an orthonormal basis $\{f_r(z)\}$
in $H_{\hbar}$ and consider the kernel
\eqn\yad{L_{\hbar}(z,{\bar z})=\sum_r f_r(z){\bar f_r(z)}.}
Functions in $A_{\hbar}$ are interpreted as symbols of operators
in $H_{\hbar}$. Multiplication in $A_{\hbar}$ is defined as
\eqn\mult{(f*g)(z,{\bar z})=
\int f(z,{\bar v}) g(v,{\bar z})G_{\hbar}(z,{\bar z}|v,{\bar v})
d\mu(v,{\bar v}),}
where the kernel $G_{\hbar}$ is
\eqn\yadr{G_{\hbar}(z,{\bar z}|v,{\bar v})=c(\hbar){L_{\hbar}(z,{\bar v})
L_{\hbar}(v,{\bar z})\over L_{\hbar}(z,{\bar z})
L_{\hbar}(v,{\bar v})}.}
It was shown in \ber\ that the correspondence principle is valid if
the following condition is satisfied
\eqn\disc{L_{\hbar}(z,{\bar z})=\lambda F^{-1/\hbar}(z,{\bar z}),}
where $\lambda$ is a constant. This condition 
restricts the Planck constant $\hbar$.
For compact complex symmetric spaces it leads to a discreteness
of $\hbar$.

There are four types of compact complex symmetric spaces
$M^I_{p,q}, M^{II}_p, M^{III}_p$ and $M^{IV}_n$.
There are special global coordinates covering
submanifolds ${\tilde M}$ in these spaces. They are given by
complex $p\times q$ matrices and also by symmetric $p\times p$,
anti-symmetric $p\times p$ matrices and by $n$-dimensional vectors
respectively. In particular $M^I_{1,1}=M^{II}_1
=S^2$ and $M^I_{1,q}$ is the complex projective space
$CP^q$.  

The \kh potential for the first three types is
\eqn\pote{F=\det(I+zz^*)^{-\nu}}
where $\nu=p+q,p+1,p-1$ for $M^I_{p,q}, M^{II}_p, M^{III}_p$
respectively. Therefore the Hilbert spaces $H_{\hbar}$
are finite dimensional and they consist of polynomials in $z$. 
The dimension of the Hilbert space $H_{\hbar}$ is
\eqn\razmer{\dim H_{\hbar}=c(\hbar)\int d\mu.}
For the correspondence principle as $\hbar \to 0$
be valid the condition \disc\ should  be satisfied
and as a result the Planck constant  takes only 
discrete values. For the space of types  I one has
\eqn\pla{{1\over \hbar}={n\over (p+q)}.}
For the space of type II
\eqn\plan{{1\over \hbar}={n\over (p+1)}}
and for the space of type III
\eqn\planc{{1\over \hbar}={n\over 2(p-1)},}
where $n=1,2,...$

In particular for the sphere $S^2$ one has
\eqn\planck{{1\over \hbar}={n\over 2}.}
The \kh potential in this case is
$F=(1+zz^*)^{-2}$ and
\eqn\dimens{\dim H_\hbar=1+1/\hbar.}
It is important to note that the discreteness 
of $1/\hbar$ is not related to the specific
quantization described above. It is proved
\bere\ that under some natural conditions
this quantization
is the unique maximum effective irreducible $w^*$
quantization up to natural equivalence.
 
Note also that if one applies the same method of quantization
to the $n$-dimensional complex space $C^n$  which is  a noncompact
\kh manifold then one obtains the standard quantum mechanics
with the infinite dimensional Hilbert space and 
without any restrictions to the Planck constant $\hbar$.

It would be interesting to use Berezin's
quantization to extend noncommutative
gauge theory \seiw\ to the case of \kh manifolds.

\newsec{Discussions}

In this note we argued that discreteness
in quantum theory on deSitter space arises because
quantum mechanics on the horizon has
a finite dimensional Hilbert space and
the coupling constant is quantized as \planck\ .

By using the AdS/CFT correspondence it was suggested in \hor\
that $EdS_4$ entropy scales as $N^2,$ where $N$
is the number of quantum degrees of freedom. Such a behavior
in principle can be consistent with the expression for the entropy
\entr\ where the Planck constant is quantized as \planck\
if $n\sim N^2$.
 We have discussed the discreteness only
in dS in 4 dimensions. It would be interesting to consider
the discreteness in various dimensions. To this end quantization
of bounded complex symmetric spaces might be suitable.
Generalization of the AdS/CFT correspondence for
the complex homogeneous domains was discussed in \bsv.

There is also another type of discreteness in the deSitter space
related to the Chern-Simons approach to $dS_3$ 
case \mal. There one has the quantization of the level in the
Chern-Simons lagrangian and it would be interesting to consider
whether this discreteness can be combined with
the discreteness discussed in this note into a unified picture.

\newsec{Acknowledgments}
I am very grateful to J. Maldacena, M. Spradlin, 
A. Strominger  and E. Witten
for helpful remarks.
This work is supported in part by DOE grant DE-FG02-91ER40654 and 
by INTAS-OPEN-97-1312.

\listrefs

\end